\documentclass[aip,pop,reprint]{revtex4-1}

\usepackage{bm}
\usepackage{graphicx}
\usepackage{epsfig}

\begin{document}

\preprint{\it The Physics of Plasmas}

\title{The Onset of Ion Heating During Magnetic
Reconnection with a Strong Guide Field} 



\author{J.~F.~Drake}
\email[]{drake@umd.edu}
\altaffiliation[permanent address: ]{University of Maryland, College Park, MD 20742}
\affiliation{University of California, Berkeley, CA 94720}


\author{M.~Swisdak}
\email[]{swisdak@umd.edu}
\affiliation{University of Maryland, College Park, MD 20742}

\date{\today}

\begin{abstract}

The onset of the acceleration of ions during magnetic reconnection is
explored via particle-in-cell simulations in the limit of a strong
ambient guide field that self-consistently and simultaneously follow
the motions of protons and $\alpha$ particles. Heating parallel to the
local magnetic field during reconnection with a guide field is
strongly reduced compared with the reconnection of anti-parallel
magnetic fields. The dominant heating of thermal ions during guide
field reconnection results from pickup behavior of ions during their
entry into reconnection exhausts and dominantly produces heating
perpendicular rather than parallel to the local magnetic field. Pickup
behavior requires that the ion transit time across the exhaust
boundary (with a transverse scale of the order of the ion sound Larmor
radius) be short compared with the ion cyclotron period. This
translates into a threshold in the strength of reconnecting magnetic
field that favors the heating of ions with high mass-to-charge. A
simulation with a broad initial current layer produces a reconnecting
system in which the amplitude of the reconnecting magnetic field just
upstream of the dissipation region increases with time as reconnection
proceeds. The sharp onset of perpendicular heating when the pickup
threshold is crossed is documented. A comparison of the time variation
of the parallel and perpendicular ion heating with that predicted
based on the strength of the reconnecting field establishes the
scaling of ion heating with ambient parameters both below and above
the pickup threshold. The relevance to observations of ion heating in
the solar corona is discussed.
\end{abstract}


\maketitle 

\section{Introduction}
\label{intro}

The production of energetic particles during flares remains a central
unsolved issue in solar physics.  Extensive observational evidence
indicates that a substantial fraction of the energy released during a
flare rapidly accelerates charged particles, with electrons reaching
$\mathcal{O}(1)$ MeV and ions $\mathcal{O}(1)$ GeV/nucleon
\cite{Emslie04}.  Explaining such energy gain requires accounting
not only for the relevant energy and time scales but also the
resulting spectra, which exhibit a common shape for most ion
species.  At the same time, high mass-to-charge ions are
greatly over-represented in flares, with abundances as much as two
orders of magnitude higher than normal coronal values
\cite{Mason94,Mason07}.

Magnetic reconnection is the ultimate energy source in impulsive
flares. Thus, in many theories reconnection plays a direct role in
particle acceleration through DC electric fields \cite{Holman85},
interactions with multiple magnetic islands \cite{Onofri06} or
first-order Fermi acceleration in contracting and merging islands
\cite{Drake06,Drake10,Oka10,Kowal11}.  On the other hand, in some models
reconnection serves as a source of magnetohydrodynamic (MHD) waves
\cite{Miller98,Petrosian04} or shocks \cite{Ellison85,Somov97} that
then independently drive particle acceleration.

The release of magnetic energy during reconnection dominantly takes
place downstream of the x-line in the exhaust where newly reconnected
field lines expand to relax their magnetic tension and drive
Alfv\'enic outflows.  The characteristic outflow speed is given by
$c_{Axup}=B_{xup}/\sqrt{4\pi m_pn_{up}}$, with $B_{xup}$ and $n_{up}$
the reconnecting component of the magnetic field and density just
upstream of the exhaust. In the MHD description the dominant heating
during reconnection of anti-parallel magnetic fields is produced by
the switch-off Petschek shocks that bound the exhaust
\cite{Petschek64}. These shocks also drive the Alfv\'enic outflow. In
the kinetic description, ions moving across the exhaust are slung by
the fast moving field lines. The resulting counterstreaming ion
distributions produce a large effective parallel temperature
$T_\parallel\sim m_ic_{Ax}^2$ with minimal perpendicular heating
\cite{Hoshino98,Gosling05}. The associated pressure anisotropy
prevents the formation of the Petschek switch-off shock
\cite{Liu12,Higashimori12}.

Coronal reconnection, however, typically involves a substantial guide
field (the magnetic field component perpendicular to the plane of
reconnection). In reconnection with a guide field the MHD model
produces a pair of rotational discontinuities (RDs) that bound the
exhaust and produce the magnetic field rotation that drives the
outflow. A pair of parallel shocks within the exhaust compress and
heat the ions.  An important question is whether the MHD description
is valid in the typical coronal environment where the particle
scattering mean-free-path is relatively long. In a kinetic description
the RDs that bound the exhaust collapse to the scale of the proton
sound Larmor radius $\rho_s=c_s/\Omega_p$, where
$c_s=\sqrt{(T_e+T_p)/m_p}$ is the sound speed and $\Omega_p$ is the
proton cyclotron frequency \cite{Drake09a}. If the crossing time of
ions (of any species) through the RD is longer than their cyclotron
time, the ions will cross the RD with little heating and will
counterstream across the exhaust to produce an increment in the
effective parallel temperature \cite{Drake09a},
\begin{equation}
\Delta T_\parallel=m_ic_{Axup}^2B_{xup}^2/B_{up}^2,
\label{Tpar}
\end{equation}
with $\Delta T_\perp \sim 0$. The ambient guide field reduces the
parallel heating compared with the case of anti-parallel reconnection.
If the crossing time of ions through the RD is shorter than their
cyclotron time, the ions entering into the exhaust are not able to
follow the rapid change in the direction of the magnetic field and
become non-adiabatic. They effectively behave like pickup particles
since they are initially at rest in the Alfv\'enic flow of the exhaust
\cite{Drake09a,Knizhnik11}. As they are ``picked up'' by the exhaust
they gain an effective thermal velocity $c_{Axup}$. Because of the
strong guide field the temperature increment is perpendicular to the
local magnetic field ($\Delta T_\parallel \sim 0$) and is given by
\begin{equation}
\Delta T_\perp =m_ic_{Axup}^2/2,
\label{Tperp} 
\end{equation}
with $m_i$ the mass of the relevant ion species -- heavier ions gain
more energy. Note that for $B_{zup}^2\gg
B_{xup}^2$ the heating in the pickup regime greatly exceeds that in
the non-pickup regime. The criterion for pickup behavior translates
into
\begin{equation}
\frac{m_i}{Z_im_p}>\left(\frac{1}{r\pi\sqrt{2}}\right)\sqrt{\beta_{xup}},
\label{massthresh}
\end{equation}
where $r=v_{in}/c_{Axup}\sim 0.1$ is the normalized rate of
reconnection, $\beta_{xup}=8\pi n_{up}(T_e+T_p)/B_{xup}^2$ is the
ratio of plasma to magnetic pressure based on the reconnection
magnetic field just upstream of the exhaust and $Z_i$ is the ion
charge state. High mass-to-charge ions satisfy this criterion more
easily than protons.

Classical collisions are often not negligible in the corona
so an important question is whether classical resistive diffusion of
the magnetic field is sufficient to broaden the RD beyond the Larmor
scale $\rho_s$. Balancing convection of magnetic flux through the RD with
resistive diffusion $\eta c^2/4\pi$ yields an equation for the
limiting width $\Delta$,
\begin{equation}
\frac{v_{in}}{\Delta}\sim\frac{\eta c^2}{4\pi\Delta^2}
\end{equation}
or
\begin{equation}
\Delta\sim\frac{\eta c^2}{4\pi rc_{Ax}}.
\end{equation}
For typical solar parameters ($T\sim 100eV$, $n\sim 10^9cm^{-3}$,
$B\sim 50G$) with $r\sim 0.1$ we find $\Delta \sim 10^{-4}cm$, much
shorter than $\rho_s\sim 30cm$. Thus, classical collisions are insufficient to broaden the RD beyond $\rho_s$ and are therefore 
unimportant. The collisionless model of the RD should correctly
describe the dynamics. 

The solar corona is normally considered a low $\beta$ medium and
therefore one might expect the inequality in Eq.~(\ref{massthresh})
for pickup behavior to be easily satisfied. However, prior to the
onset of reconnection ambient current layers are likely to be
macroscopic and for this reason when reconnection first onsets the
magnetic field just upstream of the reconnection region $B_x$ will be
very small. $B_x$ will then increase in time as the larger magnetic
field upstream convects toward the reconnection site. Thus, at the
start of reconnection all ions will be in the adiabatic regime because
$\beta_x$ will be large and energy going into ion heating compared
with that associated with the bulk flow will be small. As $\beta_x$
drops as reconnection proceeds each ion species will sequentially
(based on their mass-to-charge) move into the pickup regime. Once
protons have entered the pickup regime the fraction of released
magnetic energy going into ion heat will be comparable to that in the
bulk flow.

Consistent with this picture, observations have revealed that in the
extended solar corona, $T_{\perp} \gg T_{\parallel}$
\cite{Kohl97,Kohl98}, suggesting that magnetic reconnection is a
potential heating mechanism for the large-scale corona. Earlier there
was an assumption that the observations of $T_\perp\gg T_\parallel$ in
the corona argued in favor of ion heating by ion cyclotron waves
\cite{Cranmer03}. This assumption needs to be re-examined. Further,
if the pickup scenario for ion heating during reconnection is correct,
the lower threshold for strong ion heating for high mass-to-charge
ions might be a mechanism for the abundance enhancement of such ions
in impulsive flares \cite{Mason07} as an alternative to proposed wave
mechanisms \cite{Miller97,Petrosian04,Liu06}.

In this manuscript we explore the onset of pickup behavior and associated
strong ion heating during reconnection using a particle-in-cell (PIC)
code with an initial state with a wide current layer and a strong
ambient guide field. We include two ion species, protons and
$\alpha$ particles, so that we can separate the onset of ion
heating based on mass-to-charge. Early in the simulation the heating
of both species is weak. As reconnection develops, the upstream value
of $\beta_x$ decreases and the $\alpha$s undergo a sharp transition to
strong perpendicular heating. The threshold for this transition is
close to that given in Eq.~(\ref{massthresh}). The protons in contrast
remain below the threshold for pickup behavior and do not exhibit the
onset of strong perpendicular heating. Because the exhaust velocity of the
simulation varies over a substantial range (as $B_x$ increases
monotonically), the scaling of parallel and perpendicular heating with
the exhaust outflow velocity is obtained and compared with the
predictions in Eqs.~(\ref{Tpar}) and (\ref{Tperp}). The simulations
therefore establish the fundamental properties of ion heating during
reconnection with a guide field and lay the groundwork for
understanding ion heating in the extended corona and during
impulsive flares and in other astrophysical and laboratory systems.

\section{Numerical Simulations}
\label{simulations}

We carry out simulations using the code {\tt p3d} \cite{Zeiler02}.
Like all PIC codes, it tracks individual particles ($\approx 10^9$ in
this work) as they move through electromagnetic fields defined on a
mesh.  Unlike more traditional fluid representations (e.g., MHD), PIC
codes correctly treat small lengthscales and fast timescales, which
are particularly important for understanding the x-line and
separatrices during magnetic reconnection.

The simulated system is periodic in the $x-y$ plane, where flow into
and away from the x-line are parallel to $\mathbf{\hat{y}}$ and
$\mathbf{\hat{x}}$, respectively, and the guide magnetic field and
reconnection electric field are parallel to $\mathbf{\hat{z}}$.  The
initial magnetic field and density profiles are based on the Harris
equilibrium.  The reconnecting magnetic field is given by
$B_x=\tanh[(y-L_y/4)/w_0]- \tanh[(y-3L_y/4)/w_0]-1$, where $w_0 $ and
$L_y$ are the half-width of the current sheets and the box size in the
$\mathbf{\hat{y}}$ direction.  Particles
are distributed in a constant-density background and two current
sheets in which the density rises in order to maintain pressure
balance with the magnetic field.  We initiate reconnection with a
small perturbation that produces a single magnetic island on each
current layer.

The code is written in normalized units in which magnetic fields are
scaled to the asymptotic value of the reversed field $B_{0x}$,
densities to the value at the center of the initial current sheets
minus the uniform background density, velocities to the proton
Alfv\'en speed $c_{A}=B_{0x}/\sqrt{4\pi m_pn_0}$, times to the inverse
proton cyclotron frequency in $B_{0x}$,
$\Omega_{p0x}^{-1}=m_pc/eB_{0x}$, lengths to the proton inertial length
$d_p=c_{A}/\Omega_{p0x}$ and temperatures to $m_p c_{A}^2$.

The proton-to-electron mass ratio is taken to be $25$ in order to
minimize the difference between pertinent length scales and hence
simulate as large a domain as possible.  It has been shown
\cite{Shay98b,Hesse99,Shay07} that the rate of reconnection and
structure of the outflow exhaust do not depend on this ratio.  Since
the ion heating examined here depends only on the exhaust geometry, we
also expect our results to be insensitive to the mass ratio. The
simulation assumes $\partial/\partial z=0$, i.e. that field and
particle quantities do not vary in the out-of-plane direction, making
this a two-dimensional simulation.

In addition to protons and electrons, we also include a number density
of $4\%$ $^4$He$^{++}$ ($\alpha$) particles in the background particle
population with an initial temperature equal to that of the electrons
and protons.  This number density does not affect the reconnection
dynamics appreciably, while still providing a large sample of
particles with $m_i/m_pZ_i > 1$ that can be used to test the
scaling of the onset relation for pickup behavior given in
Eq.~(\ref{massthresh}).

\begin{figure*}
\includegraphics[width=8.5cm]{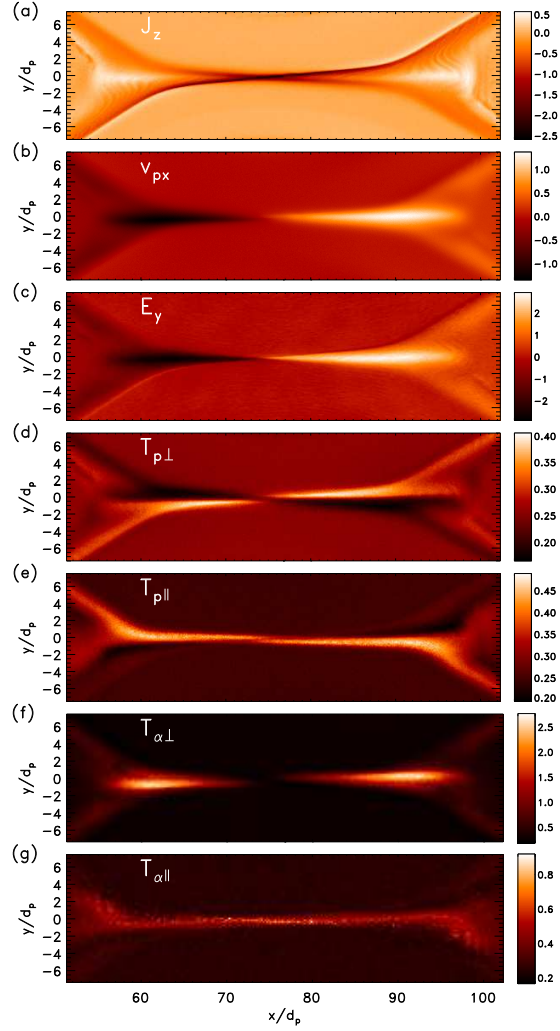}
\caption{\label{overview} (color online) Overview of a PIC simulation with an initial
  guide field $B_{0z}=2B_{0x}$ at time $t=600\Omega_{p0x}^{-1}$.  In
  (a) the total out-of-plane current density $J_{z}$; in (b) the
  proton outflow velocity $v_{px}$; in (c) the transverse electric
  field $E_y$; in (d) the proton perpendicular temperature
  $T_{p\perp}$; in (e) the proton parallel temperature
  $T_{p\parallel}$; in (f) the $\alpha$ perpendicular temperature
  $T_{\alpha\perp}$; and in (g) the $\alpha$ parallel temperature
  $T_{\alpha\parallel}$}
\end{figure*}

In Fig.~\ref{overview} we show an overview of results from a
simulation with a computational domain $L_x\times L_y=102.4 \times
51.2\,d_p$ and an initial guide field $B_{0z} = 2.0B_{0x}$ at
$t=600\Omega_{p0x}^{-1}$. The grid spacing for this run is
$0.025\,d_p$, the electron, proton, and $\alpha$ temperatures,
$T_e=T_p=T_{\alpha}=0.25m_pc_{A}^2$, are initially uniform, and the
velocity of light is $15c_{A}$. The half-width of the initial current
sheet, $w_0$, is $7\,d_p$ and the background density is $0.2n_0$.
Panel (a) depicts the total out-of-plane current density $J_{z}$
centered around the x-line of one of the current sheets.

Ambient plasma from above and below slowly flow toward the current
sheet while embedded in oppositely directed magnetic fields (pointing
to the right above the layer and to the left below).  Reconnected
field lines are bent and, to reduce their magnetic tension,
rapidly move away from the x-line, dragging plasma with them. In
Fig.~\ref{overview}(b) is the proton outflow velocity $v_{px}$. In
Fig.~\ref{overview}(c) is $E_y$, which is the electric
field that spans the exhaust during reconnection with a guide field
and forces ${\bf E}\cdot {\bf B}\sim E_yB_y+E_zB_z\sim 0$. Thus, $E_y\sim
E_zB_z/B_y$ \cite{Drake09a} and since $B_y\sim 0.1B_x\ll B_z$, $E_y$ is
much greater than the reconnection electric field $E_z$. $E_y$ controls
the outflow with $v_{px}\sim cE_y/B_z$, which is the reason for the
similarity between $v_{px}$ and $E_y$ in Fig.~\ref{overview}. Because
$E_y$ is the dominant component of ${\bf E}$, it is also the driver of
ion heating during the pickup process \cite{Drake09a}.

\begin{figure}

\includegraphics[width=8.5cm]{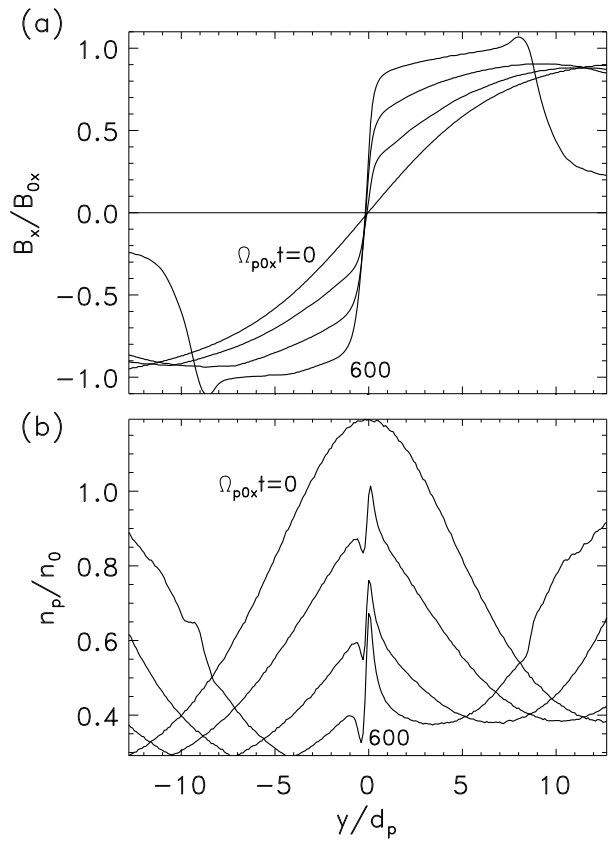}

\caption{\label{cuts} Cuts across the current layer of the
  reconnection field $B_x$ and the proton density $n$ at $t=0$,
  $t=450\Omega_{p0x}^{-1}$,$t=560\Omega_{p0x}^{-1}$ and
  $t=600\Omega_{p0x}^{-1}$.}
\end{figure}

The data shown in Fig.~\ref{overview} is at late time after the
reconnecting magnetic field $B_x$ just upstream of the x-line is
large. The time development of $B_x$ and the density $n$ in cuts
across the x-line (through $x=77d_p$ in Fig.~\ref{overview}) are shown
in Fig.~\ref{cuts} at $t=0$,
$t=450\Omega_{pi}^{-1}$,$t=560\Omega_{pi}^{-1}$ and
$t=600\Omega_{pi}^{-1}$. The current layer is initially broad with a
high density in the center of the sheet. As time passes, the magnetic
field convects inward toward the x-line at $y=0$. The magnetic field
upstream of the current layer therefore increases with time. At the
same time lower density plasma also convects toward the x-line (the
asymmetry in the cut in the density is a consequence of the density
cavities that develop during reconnection with a guide field
\cite{Pritchett04}). The consequence is that the increasing magnetic
field and lower density causes the Alfv\'en speed $c_{Axup}$ based on
the parameters just upstream of the strong current layer at the x-line
to increase dramatically. The exhaust velocity therefore also
increases with time. The relationship between the peak outflow
velocity $v_{px}$ and $c_{Axup}$ is shown in Fig.~\ref{tperppred}. The
individual data points correspond to different times as reconnection
develops.  At late time $v_{px}$ is very close to $c_{Axup}$, as
expected from the Wal\'en relation \cite{Sonnerup81}.

\section{ION PICKUP AND HEATING}\label{pickup}

Particle acceleration is controlled by the structure and magnitude of
the electric field, which, for such a strong guide field, is dominated
by $E_y$.  Particles enter the exhaust with velocity $v_y
\sim 0.1c_{Ax}$.  The non-adiabatic particles cross the boundary
in a time that is short compared with their cyclotron period and are
essentially at rest in the simulation frame while the outflow streams
past at roughly the Alfv\'en speed.  The trajectories of non-adiabatic
particles are cycloids which can be represented by an $\mathbf{E}
\boldsymbol{\times} \mathbf{B}$ drift plus an effective ``thermal
velocity'' equal to the Alfv\'en speed. The dynamics is analogous to
that of stationary neutral atoms surrounded by the moving solar wind.
When they are ionized, the new ion first moves in the direction of the motional
electric field in order to gain the necessary energy to flow with the
rest of the wind.  As it gets ``picked up'', it gains a thermal
velocity equal to the solar wind velocity \cite{Mobius85}.

In Fig.~\ref{overview} we show the temperatures of both species
parallel and perpendicular to the local magnetic field at
$t=600\Omega_{ci}^{-1}$. In (d) and (e) are the perpendicular and
parallel temperatures of the protons while in (f) and (g) are the
corresponding temperatures of the $\alpha$ particles.  The $\alpha$
temperature increase in the direction perpendicular to ${\bf B}$ is
much greater than its parallel temperature and also much greater than
that of the protons (note the different scales of the color bars). The
$\alpha$ heating is more than mass-proportional, which would be the
expected value if both species had the same thermal speed. 

\begin{figure}
\includegraphics[width=8.5cm]{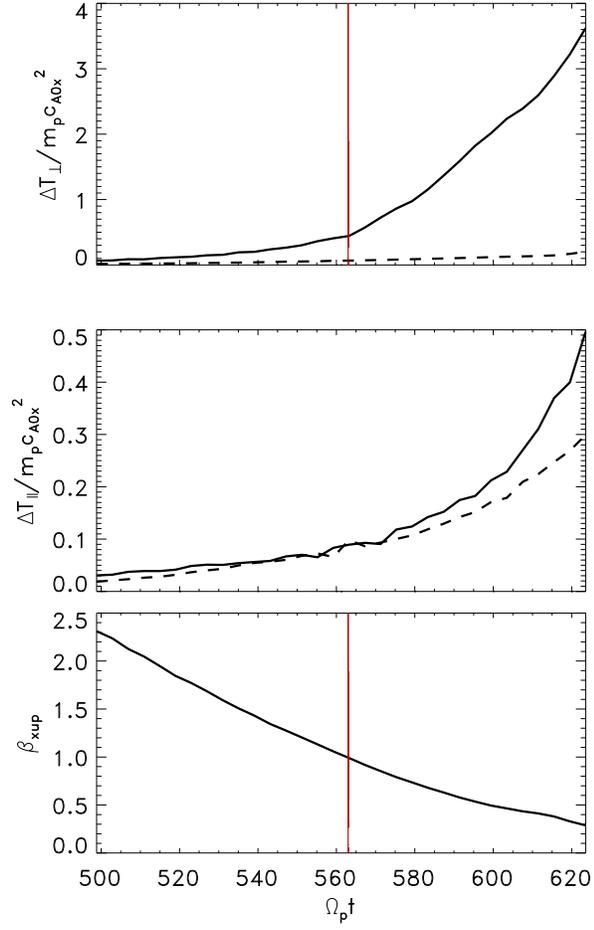}
\caption{\label{temptime} (color online) The time dependence of the increments in the proton (dashed) and $\alpha$ (solid) (a) perpendicular and (b) parallel temperatures and (c) the upstream value of $\beta_{x}=8\pi n(T_i+T_e)/B_x^2$. The vertical solid lines in (a) and (c) mark the onset of strong perpendicular heating of the $\alpha$s.}
\end{figure}

In Fig.~\ref{temptime} we show the time-development of the proton and
alpha temperature increment as well as the evolution of the upstream
value of $\beta_{xup}$. In Fig.~\ref{temptime}(a) are the increments
of the perpendicular temperatures of the $\alpha$s (solid) and protons
(dashed). There is a sharp increase in the rate of increase of the
$\alpha$ perpendicular temperature at $t=563\Omega_{p0x}^{-1}$. This
corresponds to $\beta_{xup}\sim 1$ (Fig.~\ref{temptime}(c)), which is
at a somewhat higher value than the predicted onset at
$\beta_{xup}\sim 0.79$ given in Eq.~(\ref{massthresh}. By contrast the
protons exhibit no sharp onset of perpendicular heating. Their
perpendicular heating onset should occur at $\beta_{xup}\sim 0.2$,
which is not reached by the end of the simulation. A simulation
carried out with a lower initial upstream temperature would enable us
to document the pickup onset of the protons. In Fig.~\ref{temptime}(b)
are the parallel temperature increments of the $\alpha$s (solid) and
the protons (dashed). The $\alpha$ parallel heating is well below the
perpendicular heating but is larger than expected at late time when
the $\alpha$s are in the pickup regime and the increment of the
parallel temperature should be very small. This might be due to
scattering. The spatial location of the $\alpha$ perpendicular and
parallel heating overlap (Figs.~\ref{overview}(f) and (g)) while the
peaks in the proton parallel and perpendicular heating in
Figs.~\ref{overview} do not. The proton parallel temperature increment
is modestly greater than that in the perpendicular temperature, which
is consistent with the protons remaining adiabatic.

\begin{figure}
\includegraphics[width=8.5cm]{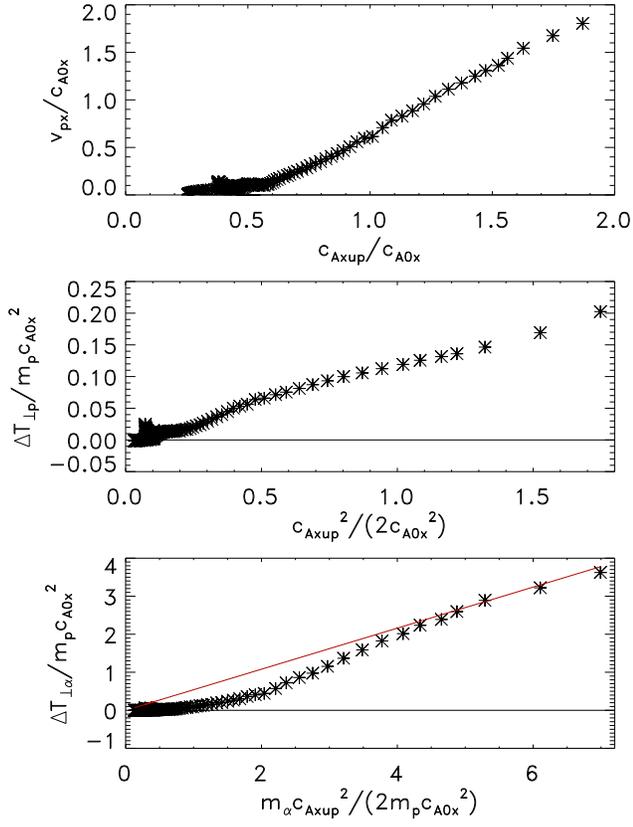}
\caption{\label{tperppred} (color online) The (a) peak proton exhaust
  velocity $v_{px}$ and the increments in the peak (b) proton and (c)
  $\alpha$ perpendicular temperatures versus their expected
  values in the pickup regime (Eq.~(\ref{Tperp})). The solid line in
  (c) has a slope of $0.54$}
\end{figure}

The ion perpendicular temperature increment is plotted versus the
expected value in the pickup regime for the protons
(Fig.~\ref{tperppred}(b)) and the $\alpha$s
(Fig.~\ref{tperppred}(c)). The proton temperature increment is far
below that which would be expected in the pickup regime (by nearly an
order of magnitude). This is consistent with the protons being
adiabatic through the end of the simulation. The $\alpha$s by contrast
have temperature increments within a factor of two of the expected
value for the larger values of $c_{Axup}$ when the $\alpha$s are in
the pickup regime. The slope of the line in Fig.~\ref{tperppred}(c) is
$0.54$ while that expected from Eq.~(\ref{Tperp}) is $1.0$. Further, that a
straight line fits through the high $c_{Axup}$
data indicatesg that $\Delta T_{\perp\alpha}\propto c_{Axup}^2$ as
expected. For the lower values of $c_{Axup}$, when the $\alpha$s are
adiabatic, $\Delta T_{\perp\alpha}$ is well below that predicted in
the pickup regime and does not scale as $c_{Axup}^2$. One reason that
the $\alpha$ perpendicular heating in the pickup regime from this
simulation is somewhat less than expected is because of a time delay
in the $\alpha$ heating relative to $v_{px}$ associated with the time
required for the $\alpha$s to be picked up by the exhaust. This is not
physically significant but is a consequence of the fact that in the
simulation at late time $v_{px}$ and $T_{\perp\alpha}$ are both
rapidly increasing (Fig.~\ref{temptime}).

\begin{figure}
\includegraphics[width=8.5cm]{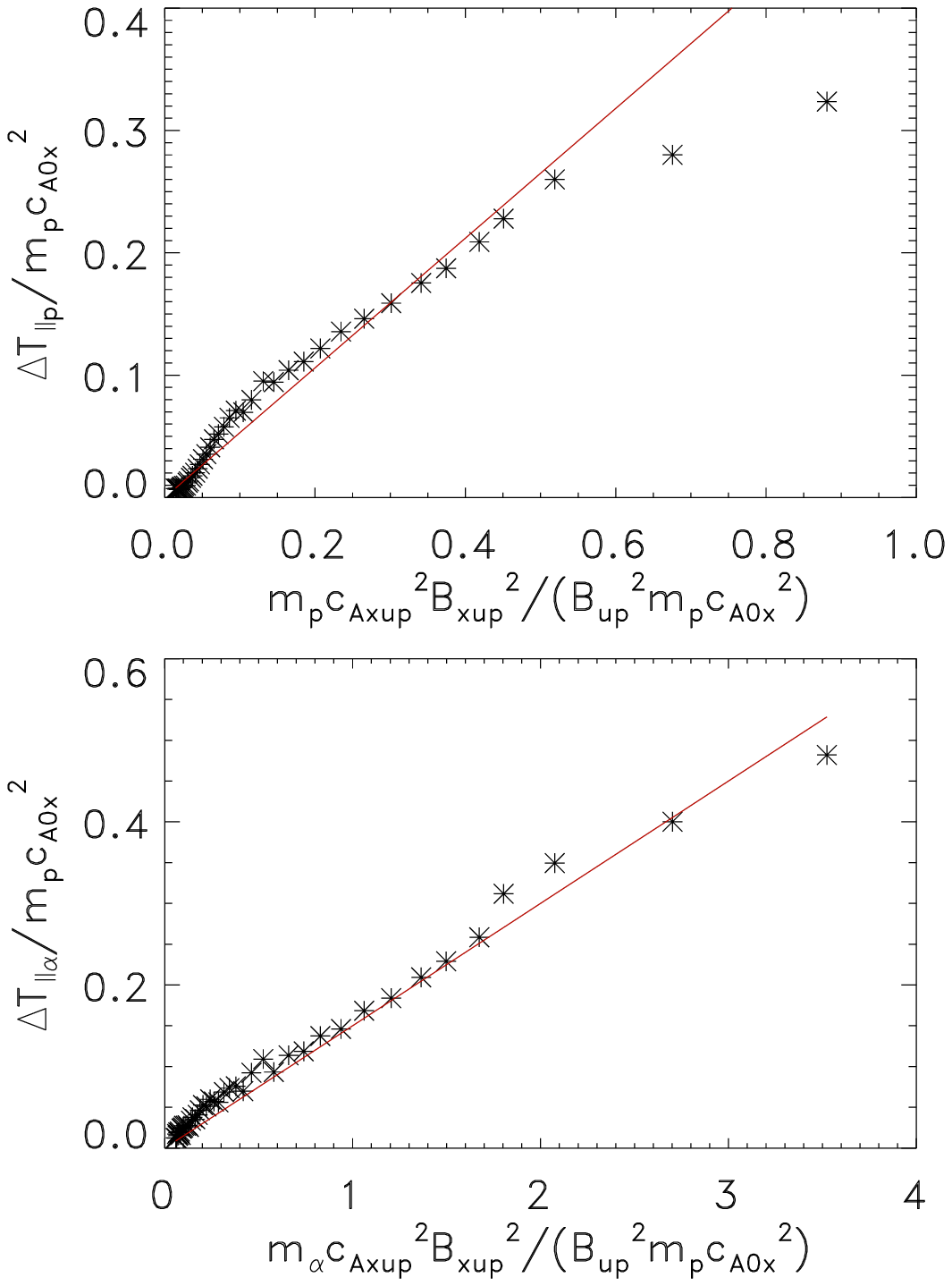}
\caption{\label{tparpred} (color online) The increments in the (a)
  proton and (b) $\alpha$ parallel temperatures versus their
  expected values in Eq.~(\ref{Tpar}) in the adiabatic regime. The solid
  line in (a) has a slope of $0.53$ and in (b) a slope of $0.15$.}
\end{figure}

Finally, in Fig.~\ref{tparpred} we show $T_\parallel$ versus the
predicted value in Eq.~(\ref{Tpar}) (the adiabatic regime) for protons
and $\alpha$s. In the adiabatic regime the perpendicular heating is
small and the ion heating falls well below the usual scaling $\Delta
T_\parallel\sim m_ic_{Ax}^2$ in the absence of a guide field. The
measured value of $\Delta T_{\parallel p}$ scales as expected but is
smaller by about a factor of two (the slope of the line in
Fig.~\ref{tparpred} is $0.54$ compared with the expected value of
$1.0$ in Eq.~(\ref{Tpar})). The data falls significantly below the
straight line at high values of $B_{xup}$, which corresponds to data
when the protons are approaching the adiabatic regime and $\Delta
T_{\parallel p}$ is expected to drop. The parallel heating of
$\alpha$s, while scaling as expected, is well below the expected value
(the slope of the line in Fig.~\ref{tparpred} is $0.15$ compared with
a slope of $1.0$ in Eq.~(\ref{Tpar})). The reason for the shortfall is
not known.
\section{DISCUSSION}

Magnetic reconnection in the corona and many laboratory experiments
typically involves a large guide field. In this regime we have shown
that ion heating at the Rotational Discontinuities (RDs) that form at
the boundaries of outflow exhausts greatly exceeds that expected from
the slow shocks of the MHD model when ions are in the pickup regime
(see Eq.~(\ref{massthresh})). For typical coronal parameters of $B =
50$G and $n = 10^9/\text{cm}^3$, energy increments of $\approx 25$
keV/nucleon are expected. The temperature increment is dominantly
perpendicular to the local magnetic field and the released magnetic
energy going into heating ions in this regime is comparable to that
associated with the bulk flow. Ions in the adiabatic regime are only
weakly heated, heating is parallel to the magnetic field and in this
regime the energy going into heating ions is much smaller than the
energy going into bulk flow.

Observations have revealed that the abundances of high mass-to-charge
ions ions are enhanced in solar flares, with the strength of the
enhancement depending only on $M/Q$. The fact that non-adiabatic
behavior and the associated strong heating depends on $M/Q$ suggests
that reconnection might be able to explain the abundance enhancements
in impulsive flares.  Furthermore, the increase in
$T_{\perp}/T_{\parallel}$ in the exhaust seen here is consistent with
that observed in the extended corona \cite{Kohl97,Kohl98}. Strong
perpendicular heating of high mass-to-charge ions is also seen in
laboratory Reversed Field Pinch experiments during global sawtooth
events \cite{Gangadhara07,Fiksel09}.

The kinetic energy of ions picked up during reconnection in the solar
corona of $\approx 25$ keV/nucleon, although significantly above
thermal energies, falls short of the inferred maximal energies of
$\approx 1$ GeV.  Further acceleration can occur via interactions with
the multiple magnetic islands predicted to be produced during flares
\cite{Drake06}.  Super-Alfv\'enic ions trapped within a slowly
contracting island can repeatedly reflect from the ends, gaining
energy via a first-order Fermi process, and producing power-law
spectra consistent with observations \cite{Drake10,Drake13}.  Thermal
ions cannot be accelerated by this process because their bounce time
is greater than the timescale for island contraction.  Thus, the
pickup process can act as a seed mechanism for further energy gain and
may ultimately control the abundances of ions measured in impulsive flares.  A
realistic test of this scenario requires 3-D simulations of multiple
x-lines and multiple islands, something which is not currently
computationally feasible. Simulations of multiple current layers in a
2-D model, however, will enable us to explore both the pickup process
and subsequent heating through island merging and contraction.

Observations of reconnection at the magnetopause
as, for example, carried out for electrons \cite{Phan13a} should yield data for 
$T_{\parallel}$ and $T_{\perp}$ for both protons and $\alpha$
particles in order to test the mechanism discussed here.  Provided the
instrumentation can differentiate between different $M/Q$ ions, data
collected by the upcoming Solar Probe Plus mission, with a planned
perihelion of $\approx 9 R_{\odot}$ (which lies within the outer
corona), should also be able put our predictions to the test.

Finally, it is widely believed that some process converts a fraction
of the energy found in the convective motions of the solar photosphere
into the heat that ensures the continuous existence of a
$\mathcal{O}(10^6\text{ K})$ corona and accelerates the solar wind.
Broadly speaking, the two most likely candidates are wave heating ---
in which oscillations generated in the photosphere travel into the
corona, develop into turbulence, and dissipate --- and reconnection,
in which the topological reorganization of the magnetic field releases
energy and heats the plasma.  Measurements by the Solar Ultraviolet
Measurements of Emitted Radiation (SUMER) and Ultraviolet Coronal
Spectrometer (UVCS) instruments of the SOHO (Solar and Heliospheric
Observatory) spacecraft provide significant constraints on any theory
of coronal heating.  In particular, at heights of $2-3 R_{\odot}$
protons have a slight temperature anisotropy with $T_{\perp} >
T_{\parallel}$ while heavier ions (represented by $O^{5+}$) are
strongly anisotropic, with $T_{\perp}/T_{\parallel} \gtrsim 10$
\cite{Kohl97,Kohl98}.  Interestingly, the process discussed in this work
should be active in the region in question and produces temperature
anisotropies consistent with these results.



%
%

%

\begin{acknowledgments}
This work has been supported by NSF Grant AGS1202330 and NASA grants
APL-975268 and NNX08AV87G. Computations were carried out at the
National Energy Research Scientific Computing Center.

\end{acknowledgments}


\end{document}